\documentclass[prb,twocolumn,superscriptaddress]{revtex4-2}
\usepackage{graphicx,color}
\usepackage{amsmath}
\usepackage{amssymb}
\usepackage{natbib}
\usepackage{tabularx}
\usepackage{xspace}
\usepackage[colorlinks=true,citecolor=blue,linkcolor=blue,urlcolor=blue,bookmarks=true]{hyperref}

\newcommand{\kbcvo}{KBa$_3$Ca$_4$Cu$_3$V$_7$O$_{28}$\xspace}

\usepackage{xspace}
\AtBeginDocument{\renewcommand{\selectlanguage}[1]{}}
\usepackage{booktabs}
\usepackage[separate-uncertainty=true]{siunitx}
\DeclareSIUnit{\eV}{eV}
\usepackage[capitalize]{cleveref}

\newcommand{\castep}{\textsc{Castep}\xspace}

\newcommand{\lsdau}{LSDA$+U$\xspace}

\newcommand{\pbeu}{PBE$+U$\xspace}

\newcommand{\cuAtom}{Cu\xspace}
\newcommand{\cuTwoPlus}{Cu$^{2+}$\xspace}
\newcommand{\vAtom}{V\xspace}

\newcommand{\ueffEq}{U_{\rm eff}}

\newcommand{\supercell}{supercell\xspace}
\newcommand{\twoBYoneBYone}{$2\times1\times1$\xspace}
\newcommand{\mBYnBYo}[3]{$#1\times#2\times#3$\xspace}
\newcommand{\monkpack}{Monkhorst--Pack\xspace}
\newcommand{\planewaveAdj}{plane-wave\xspace}

\begin{document}

\title{Quantum spin liquid on a 3D bipartite lattice of spin trimers stabilized by enhanced effective anisotropy}

\author{M.\,Gomil\v{s}ek}
\email{matjaz.gomilsek@ijs.si}
\affiliation{Jo\v{z}ef Stefan Institute, Jamova c.~39, SI-1000 Ljubljana, Slovenia}
\affiliation{Faculty of Mathematics and Physics, University of Ljubljana, Jadranska u. 19, SI-1000 Ljubljana, Slovenia}

\author{L. Mangin-Thro}
\affiliation{Institut Laue-Langevin, 71 avenue des Martyrs, CS 20156, F-38042 Grenoble, Cedex 9, France}

\author{T. Arh}
\affiliation{Jo\v{z}ef Stefan Institute, Jamova c.~39, SI-1000 Ljubljana, Slovenia}

\author{S. Petit}
\affiliation{Laboratoire L\'eon Brillouin, CEA, CNRS, Universit\'e Paris-Saclay, CE-Saclay, F-91191 Gif-sur-Yvette, France}

\author{B. Grenier}
\affiliation{Université Grenoble Alpes, CEA IRIG/MEM/MDN, F-38000 Grenoble, France}

\author{V. Simonet}
\affiliation{Institut N\'eel, CNRS--UGA, F-38042 Grenoble, France}

\author{M. Pregelj}
\affiliation{Jo\v{z}ef Stefan Institute, Jamova c.~39, SI-1000 Ljubljana, Slovenia}
\affiliation{Faculty of Mathematics and Physics, University of Ljubljana, Jadranska u. 19, SI-1000 Ljubljana, Slovenia}

\author{A.\,Zorko}
\affiliation{Jo\v{z}ef Stefan Institute, Jamova c.~39, SI-1000 Ljubljana, Slovenia}
\affiliation{Faculty of Mathematics and Physics, University of Ljubljana, Jadranska u. 19, SI-1000 Ljubljana, Slovenia}

\author{B. Koteswararao}
\affiliation{Department of Physics, Indian Institute of Technology Tirupati, Tirupati 517 619, India}

\author{B.-G. Jeon}
\affiliation{Department of Physics and Astronomy and Institute of Applied Physics, Seoul National University, Seoul 151-747, Republic of Korea}

\author{B. Sana}
\affiliation{Department of Physics, Indian Institute of Technology Madras, Chennai-600036, India}

\author{Y. Furukawa}
\affiliation{Ames National Laboratory, U.S. DOE, Ames, Iowa 50011, USA}
\affiliation{Department of Physics and Astronomy, Iowa State University, Ames, Iowa 50011, USA}

\author{Y. Inagaki}
\affiliation{Institute for the Advancement of Higher Education (Natural Sciences Section), Okayama University of Science, Okayama 700-0005, Japan}
\affiliation{Ames National Laboratory, U.S. DOE, Ames, Iowa 50011, USA}

\author{T. Asano}
\affiliation{Department of Applied Physics, University of Fukui, Fukui 910-8507, Japan}

\author{C. Repellin}
\affiliation{LPMMC, CNRS--UGA, F-38042 Grenoble, France}

\author{B. Fåk}
\affiliation{Institut Laue-Langevin, 71 avenue des Martyrs, CS 20156, F-38042 Grenoble, Cedex 9, France}

\author{J. Ollivier}
\affiliation{Institut Laue-Langevin, 71 avenue des Martyrs, CS 20156, F-38042 Grenoble, Cedex 9, France}

\author{F. Fauth}
\affiliation{ALBA Synchrotron, Carrer de la Llum 2-26 08290 Cerdanyola del Vall\`es, Barcelona, Spain}

\author{C. V. Colin}
\affiliation{Institut N\'eel, CNRS--UGA, F-38042 Grenoble, France}

\author{E. Pachoud}
\affiliation{Institut N\'eel, CNRS--UGA, F-38042 Grenoble, France}

\author{V. Pomjakushin}
\affiliation{Laboratory for Neutron Scattering and Imaging, Paul Scherrer Institute (PSI), Villigen PSI, CH-5232, Switzerland}

\author{J.S. Lord}
\affiliation{ISIS Pulsed Neutron and Muon Source, STFC Rutherford Appleton Laboratory, Chilton,OX11 0QX, UK}

\author{H. Luetkens}
\affiliation{PSI Center for Neutron and Muon Sciences, 5232 Villigen PSI, Switzerland}

\author{K.-H. Kim}
\affiliation{Department of Physics and Astronomy and Institute of Applied Physics, Seoul National University, Seoul 151-747, Republic of Korea}

\author{P. Khuntia}
\email{pkhuntia@iitm.ac.in}
\affiliation{Department of Physics, Indian Institute of Technology Madras, Chennai-600036, India}
\affiliation{Quantum Centre of Excellence for Diamond and Emergent Materials, Indian Institute of Technology Madras, Chennai 600036, India}


\begin{abstract}
Quantum spin liquids (QSLs) represent highly entangled states of matter in which frustration-induced strong quantum fluctuations suppress any symmetry-breaking phase transition down to absolute zero, and support fractionalized excitations and emergent gauge fields. Theory predicts that bond anisotropy can stabilise QSLs even on bipartite lattices, as in the celebrated Kitaev honeycomb model, but no material has so far been shown to realise such a state as the true ground state. Here we identify the three-dimensional spin-trimer magnet \kbcvo as the first bipartite QSL persisting to the lowest temperatures. Strongly coupled \cuTwoPlus trimers condense on cooling into effective pseudospins-1/2 that form a 3D bipartite network. Bulk thermodynamics, neutron scattering, $\mu$SR, and NMR detect nether  spin-freezing nor symmetry breaking phase transition down to 20~mK, but reveal a gapless dynamical ground state with algebraic spin autocorrelations. Monte Carlo and exact-diagonalisation calculations trace its stabilisation to a strong enhancement of effective anisotropy: a weak microscopic Cu--Cu anisotropy of $\sim$15\% compared to the isotropic interaction is found to generically yield effective pseudospin--pseudospin interaction anisotropies of up to 60--100\%. Trimer-based networks thus emerge as a promising platform for realizing anisotropy-stabilized quantum-entanged states even in 3D bipartite and microscopically only weakly anisotropic spin systems. 
\end{abstract}

\date{\today}

\maketitle


\section{Introduction}

The interplay between competing degrees of freedom, lattice symmetry, and quantum fluctuations can give rise to exotic states in quantum matter, including quantum spin liquids. Quantum spin liquids are highly entangled states of matter characterized by the absence of magnetic order despite strong exchange interactions, gauge field, and supports the emergence of fractionalised excitations  including spinons and Majorana  Fermions~\cite{khatua2023,Lacroix2011,Anderson1973,Wen2002,Savary2017,Zhou2017,Knolle2019,Capponi2025}. Highly frustrated magnets based on triangular motifs, such as the kagome, pyrochlore, and hyperkagome lattices and and materials with exchange frustration are prime candidates to host QSLs, as strong frustration induces strong quantum fluctuations that destabilize symmetry-breaking phase transitions down to absolute zero~\cite{Mendels2007,Hastings2000,Singh2007,Ran2007,Yan2011,Messio2012,Khuntia2020,Jiang2012,Depenbrock2012,He2017}. A complementary route, exemplified by the celebrated Kitaev honeycomb model~\cite{Kitaev2006,Jackeli2009,Kitaev2003,Nayak2008}, is to stabilise a QSL on a bipartite lattice owing to bond-dependent anisotropic interactions between pseudo spin 1/2 moments. However, despite intense efforts, no material has been conclusively demonstrated to host  Kitaev QSL on a  bipartite spin-lattice.

The experimental realization of QSLs is rare in three-dimensional (3D) spin lattices, as the enhanced connectivity of frustrated networks suppresses quantum fluctuations and favors long-range magnetic order. Notable 3D QSL candidates include the dipole--octupole pyrochlores Ce$_2$Zr$_2$O$_7$ and Ce$_2$Sn$_2$O$_7$, where Kramers-doublet Ce$^{3+}$ ions realise an octupolar $U(1)$ QSL~\cite{Gaudet2019,Gao2019,Poree2025}, and the hyperkagome Na$_4$Ir$_3$O$_8$ and PbCuTe$_2$O$_6$~\cite{Zhou2008,Khuntia2020}. Remarkably, the three-dimensional analogues of the Kitaev honeycomb model, such as the hyperhoneycomb lattice, have been proposed as promising platforms for the faithful realization of three-dimensional quantum spin liquids.  In addition, structural distortions, anisotropy, and competing exchanges can also drive 3D QSLs~\cite{Baltz2016,Alexanian2025,khatua2022signature}. 
Often, the required anisotropy originates from a crystal electric field (CEF) splitting of the energy levels of high-spin rare-earth magnetic ions, with their local pseudospin ground states inheriting strongly spin-anisotropic properties~\cite{arh2022ising}. 
An alternative, largely unexplored route uses low-spin trimer building blocks~\cite{Zhitomirsky2005,Mila1998,Repellin2017,Iqbal2020,Jahromi2020,Robert2008,Wessel2001,Fouet2006,Georgeot2010}: each $S=1/2$ trimer hosts a low-energy Kramers doublet local ground state that acts as an emergent spatially extended pseudospin-1/2. As we show here, such trimer doublets generically acquire strongly anisotropic effective inter-trimer interactions through a projection of microscopic spin--spin exchanges onto the doublet subspace, placing them in precisely the regime where an anisotropy-stabilized QSL becomes possible.In this context, the search for quantum spin liquids on trimer based 3D bipartite spin lattices, driven by competing exchange interactions and anisotropy, provides a promising platform beyond geometrically frustrated spin lattices.

Here, we present  the rare realization of a QSL on a 3D bipartite lattice based quantum magnet \kbcvo (KBCVO), characterized by persistent spin dynamics down to the lowest temperatures much below the characteristic exchnage energy scale. Strongly coupled \cuTwoPlus trimers (intra-trimer $J\approx 260$~K) condense, on cooling, into pseudospins-1/2 forming a 3D bipartite network with much weaker inter-trimer coupling $J'\sim 1$~K. Our comprehensive experimental probes, including specific heat, thermal conductivity, neutron scattering, $\mu$SR, and NMR, detect no signatures of long-range magnetic ordering or spin freezing down to 20 mK. These complementary probes reveal a gapless QSL state with algebraic spin correlations. Monte Carlo along with exact-diagonalisation (MC+ED) calculations show that projection of weakly anisotropic ($\sim$15\%) microscopic Cu--Cu exchanges onto the trimer Kramers doublets generically yields effective pseudospin anisotropies of up to 60--100\%, naturally placing the system in the effectively strongly anisotropic regime. The present work demonstrates a QSL state with short range spin correlations and exotic low-energy excitations driven by competing exchange interactions and anisotropy in a 3D bipartite trimer-based quantum magnet. 


\section{Trimers as emergent pseudospins-1/2}

\begin{figure}[!]
\includegraphics[width=1.0\columnwidth]{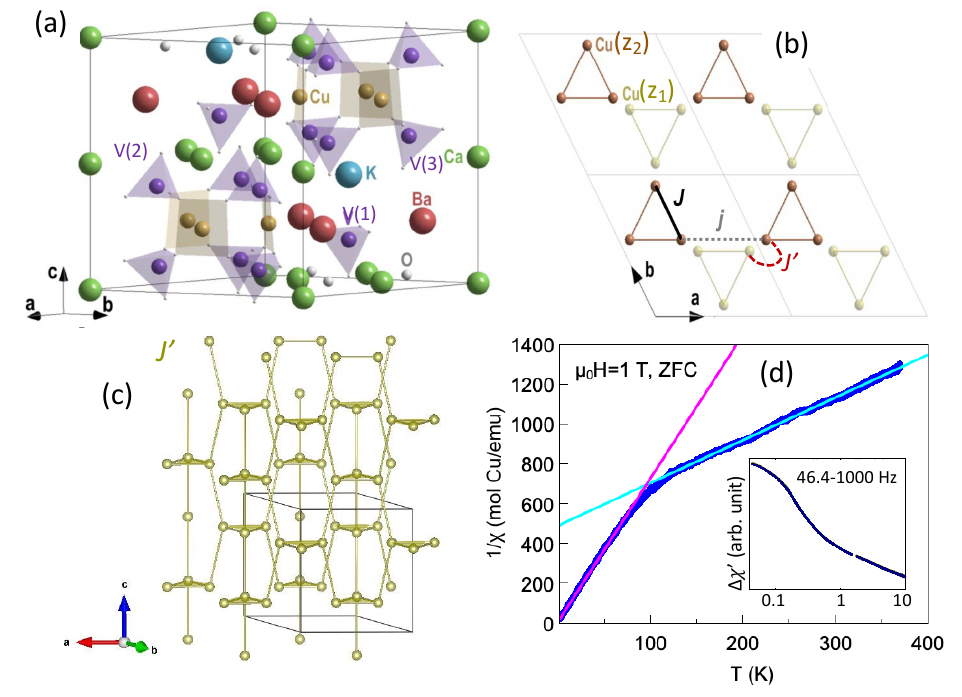}
\caption{\textbf{Crystal structure and trimer signatures in KBCVO.} \textbf{a,}~Crystal structure ($P6_3mc$, No.~186). \textbf{b,}~Cu sublattice with intra-trimer ($J$), in-plane inter-trimer ($j$) and inter-plane inter-trimer ($J'$) exchanges. \textbf{c,}~3D bipartite network defined by $J'$. \textbf{d,}~Inverse DC susceptibility $1/\chi(T)$ at $\mu_0H=1$~T (ZFC); cyan and magenta lines are Curie--Weiss fits at high and low temperatures, yielding $\theta_{\rm CW}=-230$~K and $-1.2$~K. Inset: AC susceptibility $\Delta\chi'(T)$ at 46.4--1000~Hz, showing no spin freezing down to 50~mK.}
\label{FIGURE1}
\end{figure}

KBCVO nominally crystallises in space group $P6_3mc$ (No.~186) with $a=11.1621$~\AA, $c=12.4283$~\AA, and with no detectable chemical disorder {(see Supplementary Information)}. Equilateral spin-1/2 \cuTwoPlus triangles formed by CuO$_4$ plaquettes (Fig.~\ref{FIGURE1}a,b) define a breathing-kagome basal plane (intra-triangle Cu--Cu distance 4.227~\AA, shortest in-plane inter-triangle distance 6.93~\AA, edge ratio 0.61). Adjacent layers are stacked with inverted breathing, and the shortest inter-plane Cu--Cu distance is 6.4~\AA.

DC magnetic susceptibility (Fig.~\ref{FIGURE1}d) shows no anomaly down to 1.9~K. A high-temperature Curie--Weiss fit yields $\mu_{\rm eff}=1.93~\mu_B$/Cu and $\theta_{\rm CW}=-230$~K, consistent with strong intra-triangle antiferromagnetic exchange. Below $\sim$100~K, the inverse susceptibility trend changes: the effective Curie constant drops to roughly one third of the high-temperature value, as expected if each three-Cu triangle has condensed into a local ground state with emergent pseudospin-1/2, with $\theta_{\rm CW}'=-1.2$~K then defining the inter-trimer interaction energy scale.

\begin{figure}[!]
\includegraphics[width=1.0\columnwidth]{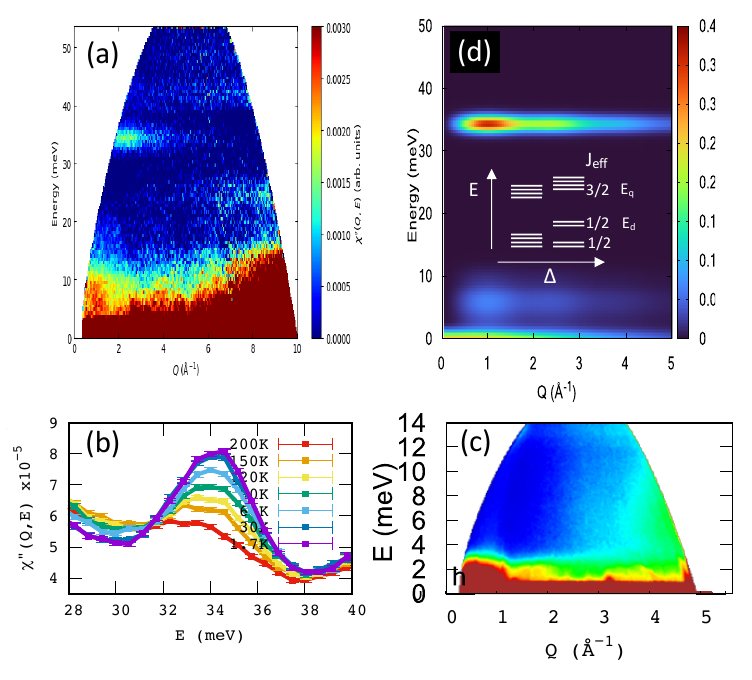}
\caption{\textbf{Trimer physics from inelastic neutron scattering.} \textbf{a,}~$S(Q,\omega)$ map (Panther, $E_i=60$~meV, $T=2$~K, high-$T$ background subtracted). \textbf{b,}~Constant-$Q$ cuts of $\chi''$ around the 34~meV mode. \textbf{c,}~IN5 map ($\lambda=2.2$~\AA, 2~K). \textbf{d,}~Calculated $S(Q,\omega)$ assuming a Gaussian distribution of intra-trimer exchanges (see Methods). Inset: trimer level scheme with a doublet--doublet splitting $\Delta=E_d$ and the upper quadruplet at $E_q$.}
\label{FIGURE2}
\end{figure}

Inelastic neutron scattering (INS) performed on the Panther instrument at ILL, France~\cite{DOI_Panther} confirms the trimer picture: a non-dispersive magnetic mode at $\sim$34~meV (Fig.~\ref{FIGURE2}) is consistent with a transition from a trimer ground state 
to an excited quadruplet at $E_q=3J/2$ with $J=263$~K (see Methods).

\textit{Ab initio} density functional theory (DFT) calculations under \lsdau and \pbeu (Methods) yield isotropic exchanges that match these experimental scales (Table~\ref{table_exchanges}). The intra-triangle $J$ overwhelms all others by roughly two orders of magnitude. Crucially, the leading inter-trimer interaction $J'$ links spins in different $ab$ planes rather than within the same plane, such that each trimer pseudospin is connected to three neighbours in a plane above and three in a plane below (Fig.~\ref{FIGURE1}c). Within numerical uncertainty, all other spin 
exchanges are at least an order of magnitude smaller. 
The relevant low-temperature lattice of pseudospins is therefore three-dimensional and \emph{bipartite}, with the shortest loops of length four. 

\begin{table}[!h]
\caption{DFT$+U$ exchanges in the high-symmetry $P6_3mc$ (No. 186) structure. Other inter-trimer spin exchanges in the supercell are bounded by the values listed; all exchanges are antiferromagnetic.}
\begin{center}
\renewcommand{\arraystretch}{1.2}
\begin{tabular}{@{} l r r r @{}}
Approach & $J$~(K) & $J'$~(K) & Other~(K) \\
\midrule
\lsdau   & $338.0(1)$         & $1.53(8)$           & ${<}\,0.10$ \\
\pbeu    & $273.4(1)$         & $1.35(5)$           & ${<}\,0.15$
\end{tabular}
\end{center}
\label{table_exchanges}
\end{table}

\begin{figure}[!]
\includegraphics[width=1.0\columnwidth]{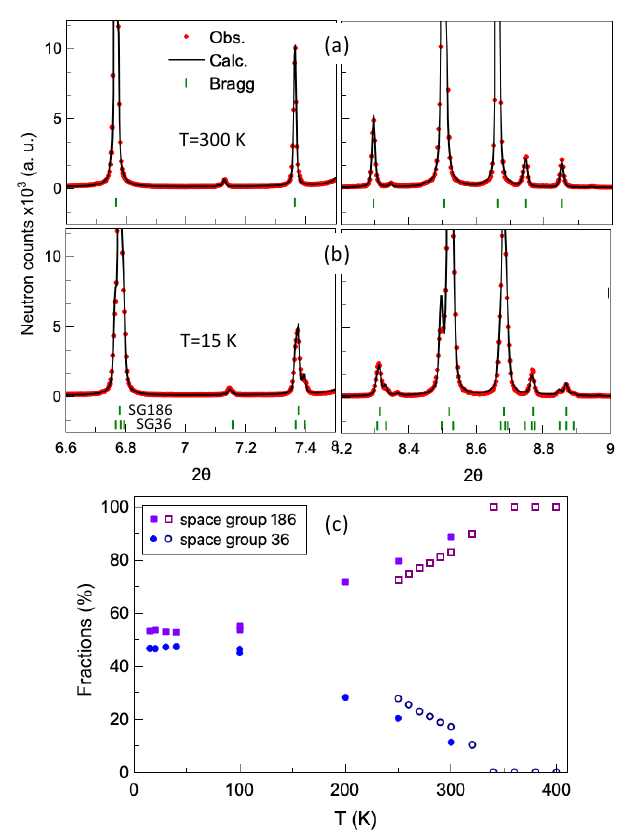}
\caption{\textbf{Structural distortion at low temperature.} Synchrotron X-ray diffraction at ALBA. \textbf{a,}~Refinement at 300~K assuming $P6_3mc$. \textbf{b,}~Refinement at 15~K with a model with a 53:47 
mixture of $P6_3mc$ and $Cmc2_1$. \textbf{c,}~Apparent phase fractions in the mixed refinement versus temperature (cryostat: filled symbols; cryofurnace: open symbols).}
\label{FIGURE3}
\end{figure}

When the constraint of the high $P6_3mc$ symmetry is lifted, \lsdau stabilises a slightly distorted structure with space group $Cc$ (No.~9; energy gain is only 0.35~eV per supercell) in which each Cu trimer is scalene: the three Cu--Cu distances in a trimer differ by $\sim$1.1\% while intra-trimer exchanges $J_{1,2,3}$ differ by $\sim$30\%, and the inter-layer $J'_{1,2,3}$ span 0.5--1.7~K (Table~\ref{table_sg9_exchanges}), consistent with the average $\theta_{\rm CW}'$. %
X-ray diffraction at the ALBA synchrotron in Spain confirms the presence of a structural distortion (Fig.~\ref{FIGURE3}): below 350~K the sample progressively converts from at most $P6_3mc$ to at most orthorhombic $Cmc2_1$ (No.~36; the \textit{ab initio} predicted space group $Cc$ is one of its maximal subgroups) symmetry, stabilising by $\sim$100~K. 
The refined variations in distorted Cu--Cu distances match the DFT predictions. 

\begin{table}[!tb]
\caption{\lsdau calculated intra-triangle ($J_i$) and inter-layer inter-trimer ($J_i'$) antiferromagnetic exchanges with the corresponding Cu--Cu distances and Cu sites (primed when inter-trimer) in the distorted structure. Other exchanges are zero within numerical uncertainty (upper bound $1.2$~K).}
\begin{center}
\renewcommand{\arraystretch}{1.2}
\begin{tabular}{@{} r r r r @{}}
Bond & Sites & Length~(\si{\angstrom}) & Strength~(K) \\
\midrule
$J_1$  & $1 \leftrightarrow 3$    & 4.251 & 191(2) \\
$J_2$  & $1 \leftrightarrow 2$    & 4.277 & 283(2) \\
$J_3$  & $2 \leftrightarrow 3$    & 4.299 & 250(2) \\
\midrule
$J_1'$ & $(1 \leftrightarrow 3)'$ & 6.299 & 1.7(1) \\
$J_2'$ & $(2 \leftrightarrow 2)'$ & 6.390 & 0.9(1) \\
$J_3'$ & $(1 \leftrightarrow 3)'$ & 6.498 & 0.5(1)
\end{tabular}
\end{center}
\label{table_sg9_exchanges}
\end{table}

For an ideal equilateral spin-1/2 triangle with isotropic interactions, its eight spin states separate into a four-fold degenerate ground state (two accidentally degenerate chiral doublets) and an excited quadruplet at $E_q=3J/2$. The structural distortion lifts this degeneracy, splitting the ground manifold into two Kramers doublets separated by $E_d\approx 3.5$~meV (inset in Fig.~\ref{FIGURE2}d; expressions in Methods). Below $T \sim E_d/k_{\rm B} \approx \SI{40}{\kelvin}$ only the lowest-energy doublet is occupied, and each trimer carries a single emergent pseudospin-1/2. 
We note that magnetoelastic coupling could broaden this mode, making it too weak to resolve in inelastic neutron scattering above the strong phonon background observed in the same energy window (\cref{FIGURE2}a,c). 
Alternatively, 
$E_d$ could be broadened if the structural transition were only partial and would result in a distribution of distorted trimer configurations. 


\section{Gapless QSL ground state on a bipartite 3D lattice}

\begin{figure}[!]
\includegraphics[width=1.0\columnwidth]{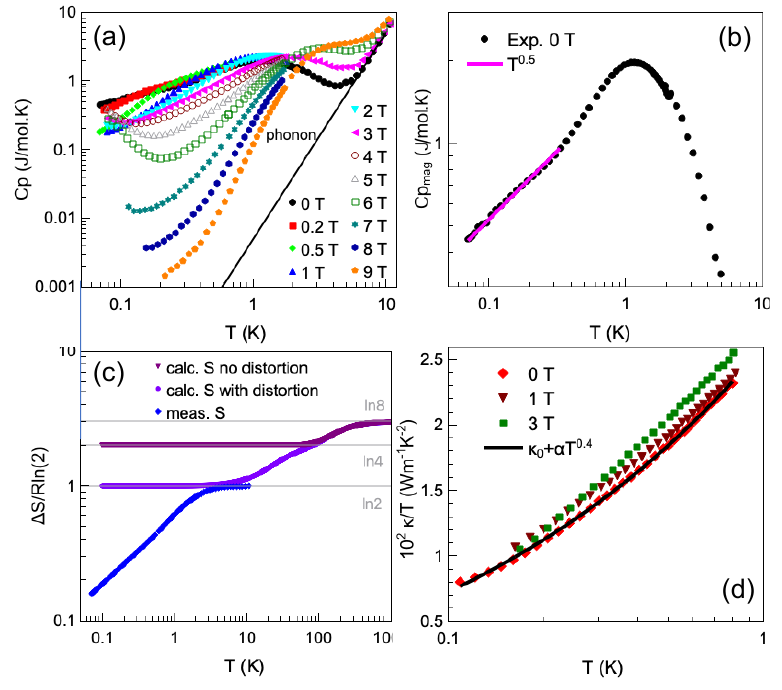}
\caption{\textbf{Bulk thermodynamic and transport probes.} \textbf{a,}~Specific heat $C_p(T)$ in 0--9~T (solid line: phonon contribution). \textbf{b,}~Magnetic specific heat $C_{\rm m}(T)$ in zero field after subtracting lattice and nuclear contributions; the magenta line is a $T^{0.5}$ fit. \textbf{c,}~Magnetic entropy per \cuTwoPlus from integration of $C_{\rm m}/T$. \textbf{d,}~Thermal conductivity $\kappa/T$ in 0, 1 and 3~T; black line is a $\kappa_0+\alpha T^{0.4}$ fit.}
\label{FIGURE4}
\end{figure}

At temperatures below $T \sim E_d/k_{\rm B}$, the inter-trimer exchange $J'$ controls the physics. AC susceptibility down to 50~mK (inset in Fig.~\ref{FIGURE1}d) shows no anomalies nor any frequency dependence, ruling out both magnetic ordering and spin freezing. The measured specific heat 
(Fig.~\ref{FIGURE4}a) shows no anomaly in zero field. After subtracting phonon and nuclear-Schottky contributions, the magnetic specific heat instead exhibits a broad maximum near 1~K and follows $C_{\rm m}\propto T^{0.5}$ below that 
(Fig.~\ref{FIGURE4}b), signalling a gapless excitation spectrum. The integrated entropy saturates at $\ln 2$ per Cu (Fig.~\ref{FIGURE4}c), consistent with a single active pseudospin degree of freedom per trimer (i.e., no undistorted triangles) and no residual entropy at zero temperature. Thermal conductivity (Fig.~\ref{FIGURE4}d) is finite and only weakly field-dependent, with $\kappa\propto T^{1.5}$ at the lowest temperatures, which provides further evidence of gapless low-energy excitations. 

\begin{figure}[!]
\includegraphics[width=1.0\columnwidth]{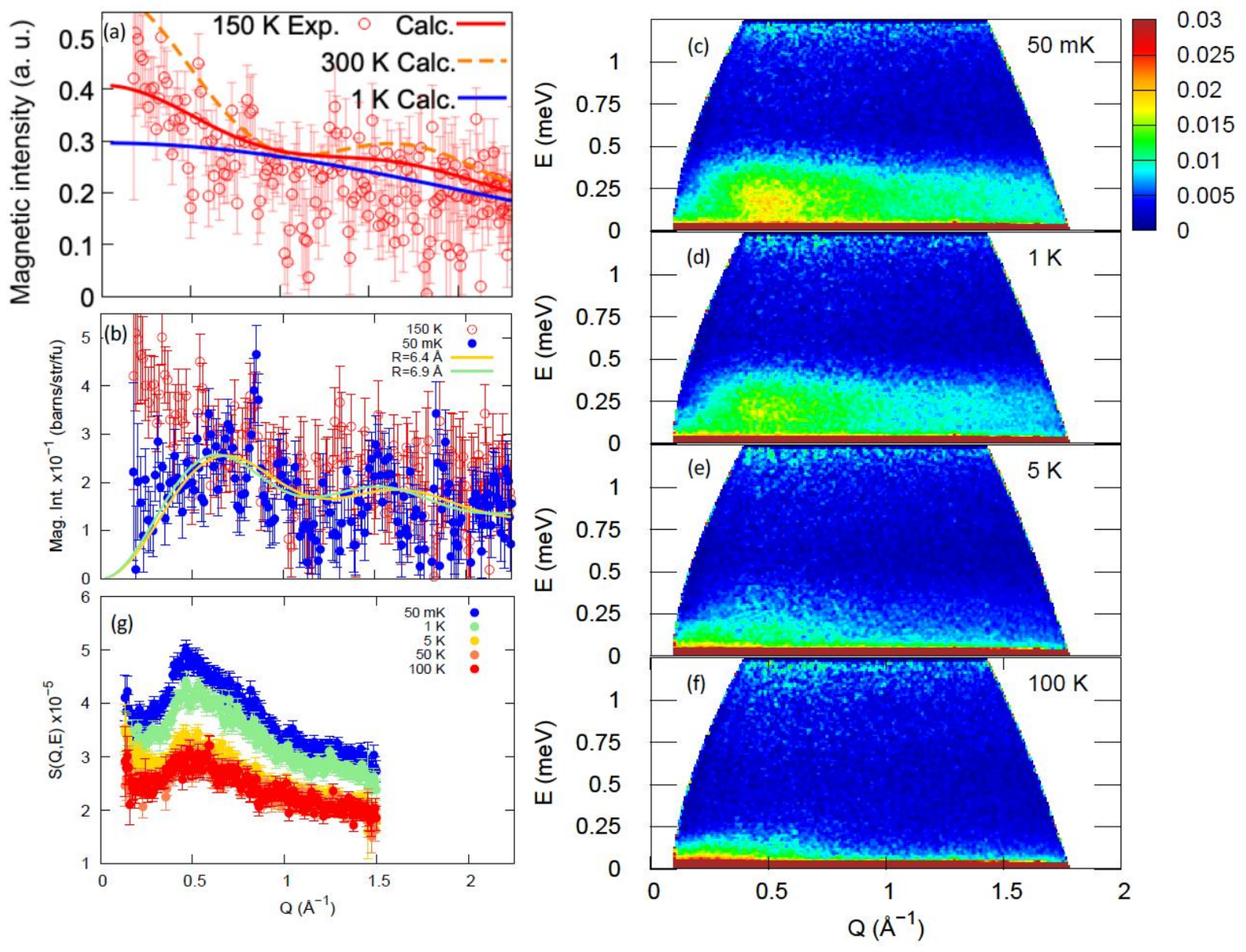}
\caption{\textbf{Magnetic correlations and gapless dynamics.} \textbf{a,}~D7 magnetic diffuse scattering at 150~K (red points) compared to exact diagonalization calculations on a distorted triangle using DFT exchanges. \textbf{b,}~D7 at 50~mK and 150~K with two-spin correlation fits (distances $R=6.4$ and $6.9$~\AA). \textbf{c--f,}~IN5 maps ($\lambda=6.5$~\AA) at 50~mK, 1, 5, and 100~K. \textbf{g,}~$Q$-cuts integrating $0.1~\mathrm{meV} < E < 1.2~\mathrm{meV}$} 
\label{FIGURE5}
\end{figure}

Polarised neutron diffuse scattering on the D7 instrument at ILL~\cite{DOI_D7a,DOI_D7b} probes the magnetic structure factor (Fig.~\ref{FIGURE5}a). At 150~K, the bump near 1.7~\AA$^{-1}$ and low-$Q$ upturn match calculated intra-trimer correlations on distorted Cu triangles assuming \textit{ab initio} exchanges, but does not match the prediction for undistorted triangles {(see Supplementary Information)}. At 50~mK, an additional broad peak develops at $Q\approx 0.7$~\AA$^{-1}$ (Fig.~\ref{FIGURE5}b), demonstrating short-range antiferromagnetic correlations between trimer pseudospins; a simple two-spin model~\cite{Bertaut,Gardner} (Methods) yields characteristic distances $R\approx 6.4$ or $6.9$~\AA, both compatible with inter-trimer separations. A density matrix renormalization group (DMRG) calculation of a purely in-plane breathing-kagome model cannot reproduce this signal {(see Supplementary Information)}, supporting the \textit{ab-initio} predicted presence of an inter-planar $J'$. Higher-resolution INS on the IN5 instrument at ILL~\cite{DOI_IN5a,DOI_IN5b} (Fig.~\ref{FIGURE5}c--g) reveals a magnetic excitation peaked near 0.2~meV and $Q\approx 0.5$~\AA$^{-1}$ that grows on cooling below $\sim$5~K and saturates at 50~mK. No spin gap is detected within the 20~$\mu$eV resolution, i.e., an order of magnitude below $J'$. Together, the finite-$Q$ diffuse scattering and gapless inelastic response are the hallmark of a 3D correlated but dynamical spin state.

\begin{figure}[!]
\includegraphics[width=\columnwidth]{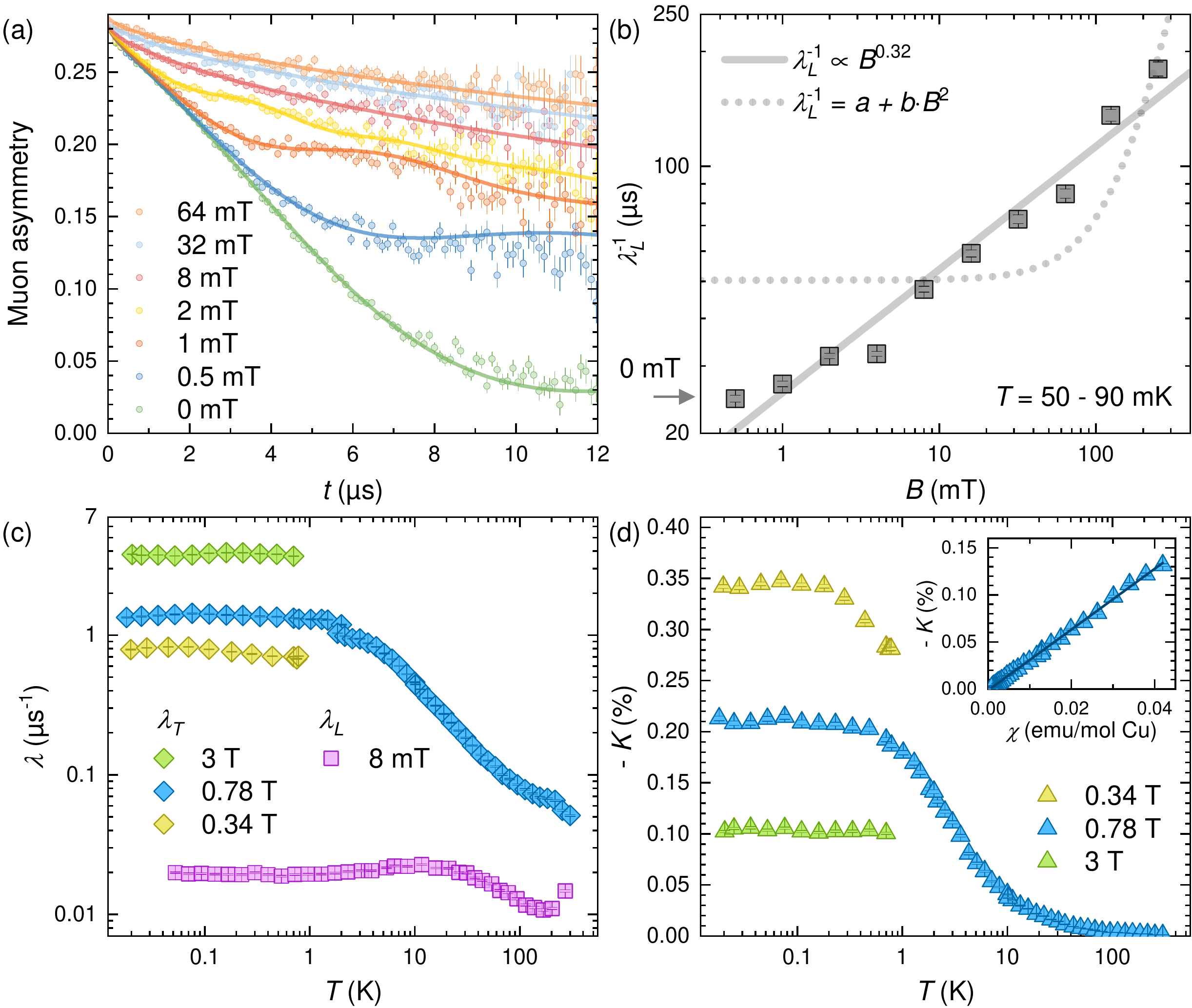}
\caption{\textbf{Persistent spin dynamics from $\mu$SR.} \textbf{a,}~Muon asymmetry at 50--90~mK in different longitudinal fields. \textbf{b,}~Field dependence of $\lambda_L^{-1}$: dotted line, Redfield fit; solid line, $\lambda_L^{-1}\propto B^{0.32}$. \textbf{c,}~Temperature dependence of $\lambda_L$ (8~mT) and $\lambda_T$ (0.34, 0.78, 3~T). \textbf{d,}~Knight shift $-K$; inset: $-K$ versus $\chi$ above 2~K.}
\label{FIGURE6}
\end{figure}

Muon spectroscopy ($\mu$SR) probes the local spin dynamics. Zero-field asymmetries (Fig.~\ref{FIGURE6}a) show no oscillations and no static $1/3$ tail~\cite{blundell2021muon}, ruling out static or long-range ordered magnetism down to at least 50~mK. A longitudinal-field decoupling experiment (Methods) yields the electronic relaxation rate $\lambda_L$. Its field dependence (Fig.~\ref{FIGURE6}b) deviates from the conventional Redfield form expected for exponentially decaying spin autocorrelations.  Instead, the data follow $\lambda_L^{-1}\propto B^p$ with $p=0.32(5)$, implying \emph{algebraic} local spin autocorrelations $S(t)\propto t^{1-p}$~\cite{gomilsek2016musr,khatua2023}, which is a non-trivial signature of strongly correlated spin dynamics, distinct from conventional paramagnets or frozen spin states. The temperature dependence of $\lambda_L$ (Fig.~\ref{FIGURE6}c) shows persistent spin dynamics in the form of a low-$T$ plateau extending from 50~mK 
to $\sim$10~K. The Knight shift $K$, accessed via transverse-field $\mu$SR, scales linearly with $\chi$ above 2~K but plateaus below 2~K in 0.34, 0.78, and 3~T (Fig.~\ref{FIGURE6}d), evidencing a finite local susceptibility down to 20~mK and confirming the gapless spectrum.

\begin{figure}[!h]
\includegraphics[width=1.0\columnwidth]{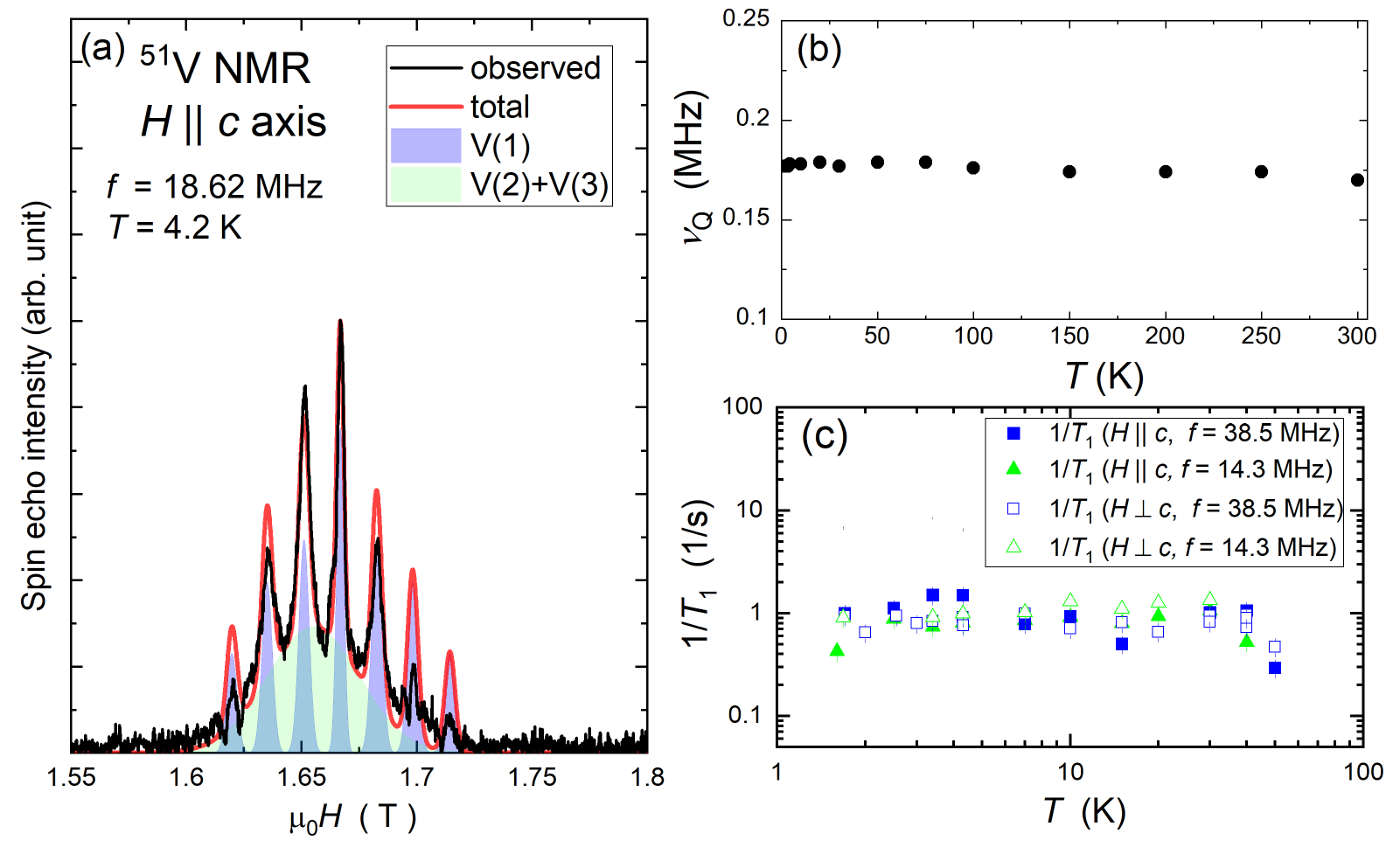}
\caption{\textbf{$^{51}$V NMR.} \textbf{a,}~Field-swept spectrum at $f=18.62$~MHz, $T=4.2$~K, $H\parallel c$; red curve is the V(1)+V(2)+V(3) sum. \textbf{b,}~$\nu_{\rm Q}(T)$ at V(1). \textbf{c,}~$1/T_1(T)$ at $f=14.3$ and 38.5~MHz, for $H\parallel c$ (filled) and $H\perp c$ (open).}
\label{FIGURE7}
\end{figure}

Finally, $^{51}$V nuclear magnetic resonance (NMR) provides a complementary microscopic view of the local spin state (Fig.~\ref{FIGURE7}). The 4.2~K field-swept spectrum is well described by a quadrupolar-split $I=7/2$ line on V(1) ($\nu_{\rm Q}=0.176$~MHz, $\eta=0$) plus a broad featureless line attributed to V(2) and V(3) sites. No magnetic anomaly is detected down to 1.6~K, and the V(1) spin--lattice relaxation rate $1/T_1$ is essentially temperature-independent in the explored window (Fig.~\ref{FIGURE7}c), which is fully consistent with the persistent low-energy fluctuations of a gapless QSL.

\section{Anisotropy enhancement stabilises the bipartite QSL}
\label{sec_anisoEnh}

The dynamical, gapless ground state observed here is unusually robust for a 3D bipartite Heisenberg antiferromagnet, which would normally order at a finite temperature. A natural mechanism for QSL stabilisation on a bipartite lattice---exemplified by the Kitaev model~\cite{Kitaev2006}---is strong bond-dependent anisotropy. We now show that, in our trimer compound, such anisotropy is expected to generically emerge from a projection of microscopic Cu--Cu exchanges onto the lowest-energy trimer Kramers doublet, even when the microscopic exchanges themselves are only weakly anisotropic.

The mechanism is the trimer analogue of CEF projection in rare-earth Kramers doublets~\cite{Gardner}: a low-energy doublet inherits exchange interactions from a microscopic Hamiltonian in a higher-dimensional Hilbert space whose matrix elements are constrained by the doublet wavefunction composition, so even small anisotropies in the microscopic spin exchanges can translate into large anisotropies in the effective inter-doublet exchanges. To quantify this for KBCVO using a Monte Carlo (MC) approach, we computed an ensemble of effective pseudospin Hamiltonians by exact diagonalisation (ED) of pairs of distorted trimers connected by the dominant inter-trimer Cu--Cu bonds $J_i'$ (\cref{table_sg9_exchanges}), 
while adding a random traceless anisotropy tensor to each $J_i$ and $J_i'$ bond of a fixed relative magnitude $\delta$ compared to the individual bond strength (full procedure in Methods).

\begin{figure}[!t]
\includegraphics[width=1.0\columnwidth]{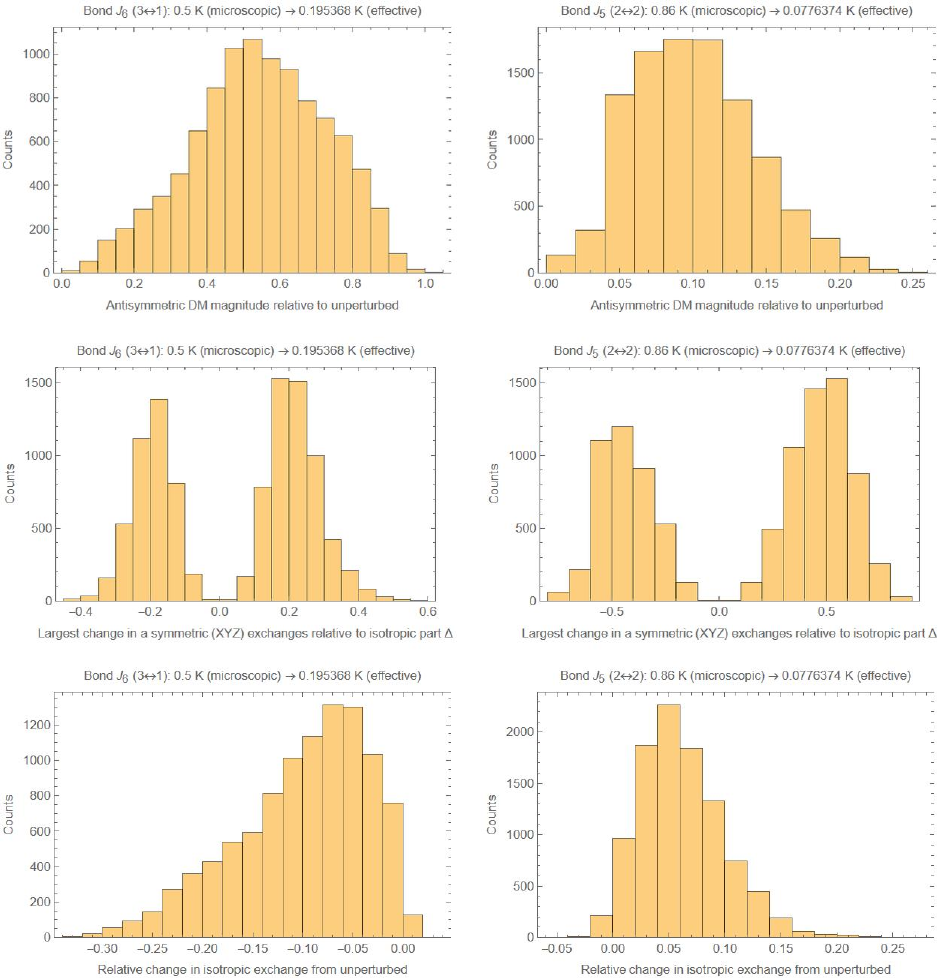}
\caption{\textbf{Projective enhancement of relative effective inter-trimer interaction anisotropy.} Histograms over $10^4$ random microscopic-anisotropy orientations of the effective bond components on $J_1'$ and $J_3'$ (left) and $J_2'$ (right), assuming a relative $\delta = \SI{15}{\percent}$ microscopic spin--spin interaction anisotropy. \textbf{Top:}~antisymmetric (Dzyaloshinskii--Moriya) magnitude relative to the unperturbed isotropic part. \textbf{Middle:}~largest symmetric (XYZ) component. \textbf{Bottom:}~relative shift of the isotropic part.}
\label{FIGURE8}
\end{figure}

Fig.~\ref{FIGURE8} shows the resulting distributions for $\delta = \SI{15}{\percent}$ microscopic anisotropy. The antisymmetric Dzyaloshinskii--Moriya (DM) component is enhanced to $\sim$60\% of the isotropic part on $J_1'$ and $J_3'$, and $\sim$10\% on $J_2'$; the symmetric XYZ-like anisotropy reaches $\sim$20\% on $J_1'$ and $J_3'$, and $\sim$50\% on $J_2'$; the isotropic part itself is renormalised by ${\sim}{-}10$\% on $J_1'$ and $J_3'$, and $\sim$10\% on $J_2'$. The mean effective magnitudes ($J_1'^{\rm eff}=J_3'^{\rm eff}\approx 0.20$ and $J_2'^{\rm eff}\approx 0.08$~K) confirm that the inter-plane bonds $J_1'$ and $J_3'$ dominate the low-energy pseudospin physics and carry the strongest anisotropy. Increasing the microscopic anisotropy to a still-realistic $\delta = \SI{25}{\percent}$ drives the mean effective DM anisotropy on $J_1'$ and $J_3'$ to $\sim$100\%; in other words, the anisotropy of the effective inter-trimer interactions 
fully equals the isotropic contribution, enabling the appearance of exotic anistropy-stabilised QSL states even on bipartite lattices. 

The projective enhancement of anisotropy in pseudospin--pseudospin interactions of the local ground states of microscopic spin clusters is generic: it is independent of the microscopic origin of the underlying spin--spin anisotropy (spin--orbit, ligand asymmetry, or lattice distortions) and does not require rare-earth spin--orbit-coupled ions. Together with the bipartite topology imposed by $J_i'$ and the gapless dynamical signatures observed in $C$, $\kappa$, INS, $\mu$SR, and NMR, this enhancement mechanism provides a coherent explanation for the QSL ground state of KBCVO.


\section{Summary and outlook}
Our results establish KBCVO as a rare 3D bipartite quantum magnet hosting a QSL state with persistent spin dynamics and short range spin correlations down to the lowest measured temperatures. Its hierarchical interaction structure with intra-trimer $J_i\approx 260$~K, inter-plane inter-trimer $J_i'\sim 1$~K, and a structural distortion that selects a single Kramers doublet per trimer naturally maps onto a 3D bipartite lattice of effective pseudospins-1/2. Our combined experimental signatures (power-law specific heat and thermal conductivity, finite-$Q$ diffuse scattering, gapless INS response, persistent spin dynamics in $\mu$SR with algebraically decaying autocorrelations, a low-$T$ Knight-shift plateau, and a temperature-independent $1/T_1$) together establish a gapless QSL state. MC+ED calculations trace its stabilisation to a projective enhancement of anisotropy in effective inter-trimer interactions.

A focused theoretical effort on the resulting bipartite pseudospin Hamiltonian, including the explicit DM and XYZ-like anisotropies extracted here, is needed to identify the precise QSL phase and its fractionalised excitations. The apparent magnetoelastic mixing in the 1--10~meV range, suggested by the absence of a clear $E_d$ doublet--doublet feature, also deserves separate study.

Finally, we show that the projective anisotropy enhancement mechanism is generic. It should operate in any quantum magnet whose elementary degrees of freedom are spin clusters with a Kramers-doublet ground state (usually, spin clusters with an odd number of half-integer-spin magnetic ions). Specifically, this mechanism establishes trimer-based spin lattices as a novel platform for stabilising anisotropy-stabilized (e.g., Kitaev-like) QSLs and other exotic spin states even in the absence of strong microscopic spin--orbit interactions, at low enough temperatures. 


\begin{acknowledgments}
First, the authors would like to have a special word of thanks to Björn Fåk, who will be sorely missed.
P.K.\ acknowledges the funding by the Anusandhan National Research Foundation (ANRF), Department of Science and Technology, India, through research grants. The work at the Ames National Laboratory was supported by Division of Materials Sciences and Engineering, Office of Science, the U.S.\ Department of Energy. The Ames National Laboratory operates at Iowa State University for the U.S.\ Department of Energy (contract no.\ DE-AC02-07CH11358). Y.~I.\ also acknowledges the JSPS Program for Fostering Globally Talented Researchers, which provided an opportunity to be a visiting scholar at Ames National Laboratory. We are grateful to large-scale facilities for beam time allocations. Experiments at the ISIS Neutron and Muon Source were supported by a beamtime allocation RB1920472 from the Science and Technology Facilities Council. Experiments at the Swiss Muon Source S$\mu$S were supported by beamtime allocations 20202615 and 20202609 from the Paul Scherrer Institute. Experiments at Institut Laue-Langevin were supported by beamtime allocations on D7 (proposal numbers 5-32-868 and 5-32-929), IN5 (proposal numbers EASY-397, 4-05-834 and EASY-1279) and Panther (proposal number 4-05-816). Experiments at ALBA were supported by beamtime allocations on BL04-MSPD (proposal number 2022025590).
\end{acknowledgments}


\section{Methods}

\subsection{Sample synthesis and characterisation}

Polycrystalline \kbcvo (KBCVO) was synthesised by a standard solid-state reaction route. Single crystals (typical size $1\times 0.75\times 0.5$~mm$^3$) were grown by flux growth using KVO$_3$ as solvent. Composition was confirmed by SEM/EDS; sample homogeneity and absence of disorder by laboratory and synchrotron X-ray diffraction. Full details are given in the {Supplementary Information}.

\subsection{Macroscopic measurements}

DC and AC magnetic susceptibility, specific heat (down to 70~mK, fields to 9~T), and thermal conductivity (down to 50~mK) were measured on single crystals; experimental details are given in the {Supplementary Information}. The lattice contribution to $C_p$ was modelled using a phonon ansatz; the nuclear-Schottky contribution arising from the applied field was subtracted following the standard procedure to obtain $C_{\rm m}$.

\subsection{Trimer Hamiltonian and level scheme}

A single trimer of three $S=1/2$ spins coupled by Heisenberg exchanges is described by ${\cal H}=J_a\,S_1\!\cdot\! S_2 + J_b\,S_2\!\cdot\! S_3 + J_c\,S_3\!\cdot\! S_1$. Defining $J=(J_a+J_b+J_c)/3$ and $j^2=(J_a J_b + J_b J_c + J_c J_a)/3$, exact diagonalisation gives a low-energy doublet pair (split by $E_d=(3/2)\sqrt{J^2-j^2}$ in the distorted case, degenerate when $J_a=J_b=J_c$) and an excited quadruplet at $E_q=E_d+3J/2$ above the ground state (Fig.~\ref{FIGURE2}d, inset). Below $k_{\rm B}T\sim J/2$ only the four lowest states matter; below $k_{\rm B}T\sim E_d$ only the lowest doublet survives, which is the emergent pseudospin-1/2. To reproduce the broadening of the 34~meV mode in Panther INS, we modelled the spectral weight using a Gaussian distribution of intra-trimer exchanges, $\Delta J=80$~K centred on $J=250$~K (Fig.~\ref{FIGURE2}d).

\subsection{Neutron and X-ray scattering}

INS data were collected at the ILL on the thermal time-of-flight spectrometer Panther ($E_i=60$~meV and 7.5~meV) and the cold time-of-flight spectrometer IN5 ($\lambda=6.5$, 8 and 2.2~\AA, down to 50~mK using a dilution insert)~\cite{DOI_IN5c}. Polarised neutron diffuse scattering with $XYZ$ polarisation analysis was measured on D7 (ILL, $\lambda=4.8$~\AA). Synchrotron X-ray diffraction was performed at ALBA (BL04-MSPD) between 13 and 400~K; Rietveld refinements used the Fullprof suite~\cite{Fullprof} and indexing of the low-temperature distorted phase was assisted by DICVOL04~\cite{DICVOL04}. Neutron diffraction at 1~K and 300~K was performed at HRPT (PSI, $\lambda=1.89$~\AA). The 50~mK D7 magnetic intensity (Fig.~\ref{FIGURE5}b) was fitted with a two-spin model $I(Q)\propto |f(Q)|^2[1-\sin(QR)/(QR)]$~\cite{Bertaut,Gardner}, where $f(Q)$ is the magnetic form factor and $R$ is the inter-trimer distance. Detailed protocols, full reductions and refinements are given in the {Supplementary Information}.

\subsection{$\mu$SR}

Zero- and longitudinal-field $\mu$SR measurements were performed on the MUSR instrument at ISIS; transverse-field measurements on GPS and HAL-9500 at PSI. Static nuclear and dynamic electronic contributions to muon depolarisation were separated by longitudinal-field decoupling, yielding the longitudinal electronic relaxation rate $\lambda_L$. The conventional Redfield form $\lambda_L^{-1}=a+bB^2$ corresponds to exponentially decaying spin autocorrelations $S(t)\propto e^{-t/\tau}$. For KBCVO, fits with this form fail (dotted line in Fig.~\ref{FIGURE6}b); the data instead follow $\lambda_L^{-1}=bB^p$ with $p=0.32(5)$ and $b=25(6)$~$\mu$s/mT$^{0.32}$ (with $a=0$ fixed), which corresponds to algebraic local-field autocorrelations $S(t)\propto t^{-(1-p)}=t^{-0.68}$. Detailed modelling of all $\mu$SR observables is given in the {Supplementary Information}.

\subsection{NMR}

$^{51}$V NMR ($\gamma_N/2\pi=11.193$~MHz/T, $I=7/2$) was performed at Ames National Laboratory on aligned single crystals. Spectra were field-swept; $T_1$ was extracted from the magnetisation recovery curve, fitted with a long+short two-component model that accounts for the overlap with the V(2)/V(3) signal {(see Supplementary Information)}.

\subsection{DFT calculations}

Total-energy DFT calculations used \castep with ultrasoft pseudopotentials, under both \lsdau ($\ueffEq=\SI{6}{\milli\electronvolt}$ for \cuAtom, consistent with values established for other copper oxides~\cite{janson2008modified,jeschke2013first,jeschke2015barlowite,iqbal2015paramagnetism,arh2020origin}; $\SI{3.5}{\milli\electronvolt}$ for \vAtom, consistent with values established for VO$_2$~\cite{stahl2020critical,cui2018first}) and \pbeu schemes~\cite{clark2005first,anisimov1997first}, on a geometry-optimised \twoBYoneBYone \supercell with a $\SI{2000}{\electronvolt}$ \planewaveAdj energy cutoff and a \mBYnBYo{1}{2}{2} \monkpack $k$-mesh~\cite{monkhorst1976special}. Heisenberg exchange constants were obtained by least-squares fitting of energies of more than 120 collinear spin configurations~\cite{riedl2019ab}. Allowing the structure to relax without imposing the $P6_3mc$ symmetry, \lsdau stabilises a slightly distorted structure of lower symmetry (space group No.~9, energy gain 0.35~eV per supercell), yielding the scalene-trimer exchanges of Table~\ref{table_sg9_exchanges}.

\subsection{Effective pseudospin Hamiltonian and anisotropy enhancement}

The effective pseudospin Hamiltonian was obtained by exact diagonalisation of pairs of distorted trimers connected by inter-trimer bonds. Microscopic Cu--Cu exchange tensors were constructed by adding to each isotropic exchange (DFT-derived) random anisotropic components corresponding to a fixed total anisotropy fraction (15\% or 25\% of the isotropic part), with random orientation. For each random configuration, we exact-diagonalised the trimer pair, projected the inter-trimer interaction onto the product of trimer ground-state Kramers doublets, and decomposed the resulting effective bond Hamiltonian into isotropic, antisymmetric DM, and symmetric off-diagonal XYZ-like components. Distributions over $10^4$ random configurations are shown in Fig.~\ref{FIGURE8}. 


\bibliography{biblio-common}
\end{document}